\documentclass[superscriptaddress,letterpaper, amssymb,preprint, amsmath, 12pt]{revtex4-1}
\usepackage[utf8]{inputenc}
\usepackage[english]{babel}
\usepackage{subcaption}
\usepackage{mathtools}
\usepackage{braket}
\usepackage{float}
\usepackage{graphicx}
\usepackage{siunitx}
\usepackage{comment}
\usepackage{url}
\usepackage[colorlinks = true,
            linkcolor = blue,
            urlcolor  = blue,
            citecolor = blue,
            anchorcolor = blue]{hyperref}

\usepackage[justification=raggedright,singlelinecheck=false,labelfont=bf]{caption}  
\newcommand{\etal}{\textit{et al.~}}
\usepackage{hyperref}
\hypersetup{
    colorlinks=true,
}

\usepackage{scalerel}
\usepackage{tikz}
\usetikzlibrary{svg.path}
\definecolor{orcidlogocol}{HTML}{A6CE39}
\tikzset{
  orcidlogo/.pic={
    \fill[orcidlogocol] svg{M256,128c0,70.7-57.3,128-128,128C57.3,256,0,198.7,0,128C0,57.3,57.3,0,128,0C198.7,0,256,57.3,256,128z};
    \fill[white] svg{M86.3,186.2H70.9V79.1h15.4v48.4V186.2z}
                 svg{M108.9,79.1h41.6c39.6,0,57,28.3,57,53.6c0,27.5-21.5,53.6-56.8,53.6h-41.8V79.1z M124.3,172.4h24.5c34.9,0,42.9-26.5,42.9-39.7c0-21.5-13.7-39.7-43.7-39.7h-23.7V172.4z}
                 svg{M88.7,56.8c0,5.5-4.5,10.1-10.1,10.1c-5.6,0-10.1-4.6-10.1-10.1c0-5.6,4.5-10.1,10.1-10.1C84.2,46.7,88.7,51.3,88.7,56.8z};}}
\newcommand\orcidicon[1]{\href{https://orcid.org/#1}{\mbox{\scalerel*{
\begin{tikzpicture}[yscale=-1,transform shape]
\pic{orcidlogo};
\end{tikzpicture}
}{|}}}}

\begin{document}

\title{Theory of drain noise in high electron mobility transistors based on real-space transfer}

\author{Iretomiwa Esho \orcidicon{0000-0002-3746-6571}}
\affiliation{%
Division of Engineering and Applied Science, California Institute of Technology, Pasadena, California 91125, USA
}

\author{Alexander Y. Choi \orcidicon{0000-0003-2006-168X}}
\affiliation{%
Division of Engineering and Applied Science, California Institute of Technology, Pasadena, California 91125, USA
}

\author{Austin J. Minnich \orcidicon{0000-0002-9671-9540}}
 \email{aminnich@caltech.edu}
\affiliation{%
Division of Engineering and Applied Science, California Institute of Technology, Pasadena, California 91125, USA
}%

\date{\today}

\begin{abstract}
 High electron mobility transistors  are widely used as  microwave amplifiers owing to their low microwave noise figure. Electronic noise in these devices is typically modeled by noise sources at the gate and drain. While consensus exists regarding the origin of the gate noise, that of drain noise is  a topic of debate. Here, we report a theory of drain noise as a type of partition noise arising from real-space transfer of hot electrons from the channel to the barrier. The theory accounts for the magnitude and dependencies of the drain temperature and suggests strategies to realize devices with lower noise figure.
 
\end{abstract}
\maketitle

\section{Introduction}

Low noise microwave amplifiers based on high electron mobility transistors (HEMTs) are widely used in scientific applications ranging from radio astronomy \cite{pospieszalski2005extremely} to quantum computing \cite{bardin2021microwaves, krantz2019quantum}. Decades of progress in device fabrication have yielded significant advances in  figures of merit such as transconductance \cite{enoki19960, saranovac2017pt, leuther200835}, gain \cite{wadefalk2003cryogenic}, unity gain cutoff frequency \cite{lai2010sub,cha20180, leuther2009metamorphic}, maximum oscillation frequency \cite{lai2007sub}, and power consumption \cite{liu2011inp, cha20180lp}.
The resulting devices exhibit excellent noise performance, with minimum reported noise figures of HEMTs around a factor of 5 above the standard quantum limit in the 1-100 GHz frequency range \cite{wadefalk2003cryogenic, cha20180, schleeh2013cryogenic, cuadrado2017broadband}.


Further improvements in the noise performance of HEMTs require a physical understanding of the microscopic origin of electronic noise. The Pospieszalski model \cite{pospieszalski1989modeling} describes the noise using noise generators at the gate and drain. The gate noise is generally attributed to thermal noise of the gate metal \cite{pospieszalski2005extremely, bautista2007physical}, but the physical origin of the drain noise remains unclear.  The earliest treatment of electronic noise in HEMTs by Pucel \etal \cite{statz1974noise} described noise in the saturated region as originating from the generation of dipole layers. More recently, drain noise has been attributed to a suppressed shot noise mechanism \cite{pospieszalski2017limits}. Experimentally, drain noise is reported to exhibit a dependence on drain current \cite{pospieszalski2005extremely} and  physical temperature \cite{murti2000temperature, weinreb2021low,munoz1997drain}, although the temperature dependence is disputed \cite{pospieszalski2017dependence}. Recent work has reported the dependence of drain noise on drain current as well as drain voltage, with the former being dominant but the latter being non-negligible in devices with  35 nm gate length \cite{heinz2020noise}.

A separate body of literature has extensively investigated high-field transport \cite{hopfel1985hot, shah1984hot, asche2006hot}, energy relaxation \cite{shah1985energy,kash1985picosecond, lyon1986spectroscopy, shah1978hot, hopfel1986nonequilibrium, knox1986femtosecond, shah1981investigation}, microwave noise \cite{aninkevivcius1993comparative, matulionis1997qw}, and related properties in 2D quantum wells \cite{hopfel1985electron, shorthose1989phonon, shah1986hot, hopfel1986picosecond, ridley1991hot}. The physical picture of high-field transport obtained from these studies is that electrons are heated by the electric field and lose energy primarily by optical phonon emission. Photoluminescence experiments provide evidence that electrons heated by the field scatter rapidly enough with each other to maintain a distribution characterized by a temperature that is higher than the lattice temperature \cite{shah1984hot}. If the electron temperature is sufficiently high, electrons may  thermionically emit over the confining potential at the heterointerface between channel and barrier and thereby leave the channel in a process known as real-space transfer (RST). This process was originally proposed as a means to realize heterostructure devices exhibiting negative differential resistance (NDR), where the NDR originates from an increased electron population in the lower mobility barrier layer as the drain voltage is increased \cite{hess1979negative}. Devices exploiting the effect, such as charge injection transistors \cite{luryi1984charge} and negative resistance field effect transistors \cite{kastalsky1984field}, were reported shortly thereafter. RST has also been observed in HEMTs under forward gate bias and high drain voltage \cite{chen1988observation, laskar1992experimental}.

Observing NDR in a HEMT requires a non-negligible fraction of channel electrons to emit into the low-mobility barrier layer.  However,  even if RST is not evident in current-voltage characteristics, it may contribute to microwave noise as a type of partition noise between two dissimilar current paths, similar to intervalley noise \cite{price1960intervalley, shockley1966quantum}.  Microwave noise in semiconductor quantum wells and devices has been previously attributed to RST. For instance, Aninkevicius \etal concluded that RST was the origin of noise in an AlGaAs/GaAs heterostructure at 80 K based on the measured dependencies of noise temperature on electric field and conduction band offset, and they further attributed intervalley noise suppression to RST at high fields \cite{aninkevivcius1993comparative}. In HEMTs, Feng \etal attributed drain noise partially to RST \cite{feng1997real}, although evidence supporting the claim was not provided. Other works reported on a RST process dominating low frequency noise characteristics of AlGaAs/InGaAs HEMTs \cite{van1993thermally}. Monte Carlo simulations have reported RST to affect the transit time \cite{mateos1996influence} and contribute to gate noise \cite{mateos1998influence}. Despite these prior studies in which noise in HEMTs was attributed to RST, a systematic examination of whether RST can account for the reported magnitude and trend of microwave drain noise in the context of the Pospieszalski model is lacking.



Here, we report an analytical theory of drain noise in HEMTs based on microwave partition noise arising from real-space transfer. The theory yields an expression for the drain noise temperature of the Pospieszalski model in which the peak electron temperature and the conduction band offset are key parameters.  The theory explains the reported dependencies of the drain temperature and makes predictions about how to reduce its magnitude. Our work may guide the development of HEMTs with improved noise performance.

\section{Theory}

Consider a two-dimensional electron gas (2DEG) with an applied longitudinal electric field  between the source and drain contacts such that electrons flow from  source  to  drain. We may focus only on the region under the gate by incorporating the other regions as access resistances \cite{schwierz2003modern}. At the low-noise bias $V_{GS} \sim -0.1$ V, the channel is pinched off, leading to an electric field with peak value $\sim 100$ kV cm$^{-1}$ \cite{mateos1999effect} under the drain side of the gate to maintain the current of around tens of \si{\milli \ampere \per \milli \meter} \cite{schleeh2013cryogenic}. The electric field heats the electrons to a temperature that may be sufficient for electrons to thermionically emit out of the channel; if so, current will flow  through both the channel and the barrier to the drain contact. The barrier is typically of much lower mobility than the channel owing to ionized impurity scattering by the dopants, and therefore NDR will result from RST if a sufficiently large fraction of electrons transfer to the barrier. 

Even if no changes in the $I_{DS}-V_{DS}$ characteristics due to RST can be detected, non-negligible current noise may still be generated by RST. The generated noise can be viewed as a type of partition noise owing to the different mobilities of the channel and barrier. As given in Eq.~4.21 of Ref.~\cite{asche2006hot}, the spectral noise power of this mechanism can be expressed in terms of frequency $\omega$ and electric field $\mathcal E$ by:

\begin{equation}
    S_j(\omega, \mathcal E) = \frac{4e^2 n_1 n_2(v_{d1} - v_{d2})^2 \tau}{V_{0}n(1+\omega \tau)^2}
    \label{Sjw}
\end{equation}

\noindent
where $\tau$ is the characteristic time for electrons to transfer from channel to barrier;  $n_1$, $n_2$, $v_{d1}$, and $v_{d2}$ are average carrier concentrations and velocity in the channel (index 1) and barrier (index 2), respectively; $n= n_1 + n_2$; and $V_0$ is the 2DEG volume.

Equation~\ref{Sjw} can be simplified further with the following considerations. First, we take $n_2 \ll n_1$ because NDR is not observed at the low-noise bias, constraining the maximum magnitude of $n_2$. Second, $v_{d2} \ll v_{d1}$ since the spacer mobility is much less than the the channel mobility.
Finally, $\omega \tau \ll 1$ in microwave applications so that Eq.~\ref{Sjw} becomes:

\begin{equation}
    S_j(\mathcal E) = \frac{4e^2 n_2 v_{d1}^2 \tau}{V_{0}}
    \label{Sj}
\end{equation}

Let $V_0 = L W d$, where  $W$ and $d$ are the gate width and 2DEG thickness, respectively, and $L$ is a characteristic length over which electrons are hot enough to undergo RST. To facilitate comparison with the Pospieszalski model, we note that the spectral density of current fluctuations ($S_I$) is related to that of current density fluctuations  ($S_j$) as $ S_I = A^2 S_j$ where $A=W d$. Then, $S_I$ can be expressed as:

\begin{equation}
    S_I(\mathcal E) = \frac{4e^2 n_{s2} v_{d1}^2 W \tau}{L}
    \label{SI2}
\end{equation}
where $n_{s2} = d n_2$ is the barrier sheet density and we have assumed that $d$ is on the order of the barrier thickness. 

This partition noise is added at the output of the HEMT. In the Pospieszalski model, the output spectral noise power $S_I =  4 k_B T_d g_{DS}$ is parametrized by a drain temperature, $T_d$, of the drain conductance $g_{DS}$. To connect Eq.~\ref{SI2} to the Pospieszalski model, we equate the spectral noise powers and solve for $T_d$. A simple expression for the drain temperature can then be obtained as:

\begin{equation}
    T_d = \frac{e^2 v_{d1}^2 n_{s2}\tau}{k_B g_{DS}' L}
    \label{Td}
\end{equation}

where $g_{DS}' = g_{DS}/W$ is the drain conductance per width. From Eq.~\ref{Td}, $T_d$ is observed to depend on $n_{s2}$, showing a direct relationship between the fraction of electrons transferred into the barrier and $T_d$. $n_{s2}$ in turn depends on the electron temperature, the conduction band offset between channel and barrier, and the probability for a hot electron to emit out of the channel.

\section{Results} \label{results}

\begin{figure}
    {
    \includegraphics[width=\textwidth]{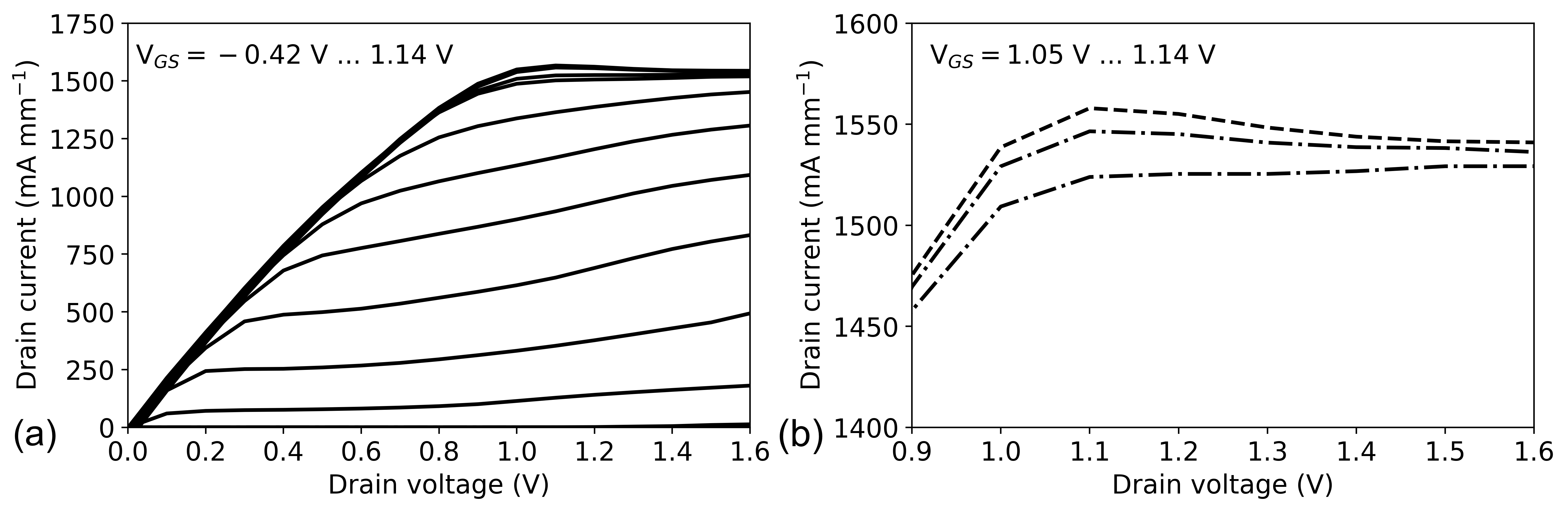}
    \phantomsubcaption\label{ndr} \phantomsubcaption\label{ndrmag}
    } \caption{(a) $I_{DS}-V_{DS}$ characteristics of a 2 $\times$ 100 $\mu$m gate width, 150 nm gate length InP HEMT at 5 K for several $V_{GS}$.  (b) Magnified view of current-voltage characteristics in (a) under forward bias. NDR is observed for $V_{GS} > 1.05$ V, consistent with the occurrence of RST.
    Data courtesy of Junjie Li and Jan Grahn, Chalmers University of Technology.
    }
    \label{IV}
\end{figure}

We assess the validity of the theory by first presenting experimental evidence of the RST process in modern HEMTs at cryogenic temperatures.  Current-voltage characteristics of an InP HEMT studied in Ref.~\cite{cha20180} were measured at 5 K. The data, courtesy of Junjie Li and Jan  Grahn,  are shown in Figure \ref{ndr} for $-0.42$ V $\leq V_{GS} \leq$ 1.14 V. Typical $I-V$ curves are observed for most values of $V_{GS}$, including those corresponding to depletion ($V_{GS} \lesssim -0.1$ V) at the low noise bias. In particular, a positive output conductance, $g_{DS} > 0$, is observed for $V_{GS} < 1$ V.

As $V_{GS}$ increases above 1 V, $g_{DS} < 0$ is observed. A magnified view of these characteristics is shown in  Fig.~\ref{ndrmag}, in which NDR is clearly present when  $V_{DS} > 1.1$ V.  This observation is consistent with RST because $V_{GS}$ controls the fraction of electrons that can emit out of the channel into the barrier. As $V_{GS}$ is increased, an increasing number of electrons are energetic enough to undergo RST, leading to a decrease in current and observable NDR. This experimental result provides evidence that RST may occur in modern HEMTs, implying that it may also produce partition noise  and that $V_{GS}$ controls the fraction of electrons in the channel that can undergo RST.

We now  examine the predictions of the theory and how they compare to the reported magnitude and dependencies of drain noise. First, to estimate the magnitude of $T_d$ from Eq.~\ref{Td}, we must specify numerical values of the various parameters. From Ref.~\cite{schleeh2013cryogenic}, typical parameters for an InAlAs/InGaAs HEMT at the low noise bias and cryogenic temperatures are obtained as $g_{DS}' \sim 50$ \si{\milli \siemens \per \milli \meter}, $L \sim 100$ nm, and $v_{d1} \sim 2 \times 10^7$ \si{\centi \meter \per \second}. The channel-barrier transit time, $\tau$, has been estimated to be on the order of picoseconds by analyzing current reduction in a test structure devised to measure switching and storage effects in GaAs/AlGaAs heterojunctions \cite{keever1982fast}. Following Eq.~5.15 in Ref.~\cite{ridley1991hot}, we choose $\tau \sim 1$ ps as a characteristic time for the emission process.

Next, the sheet density in the barrier due to transferred electrons, $n_{s2}$, is required. This parameter depends on the channel sheet density in the pinched off region under the gate $n_{s1}$; the hot electron fraction $\eta$, or the fraction of electrons that are energetic enough to thermionically emit over the barrier; and the probability for a hot electron to actually emit, $\gamma$; as $n_{s2} \equiv \gamma \eta n_{s1}$. For simplicity, we assume that all electrons with sufficient energy jump the barrier so that $\gamma = 1$. Next, we find $n_{s1} \sim 10^{11}$ \si{\per \centi \meter \squared} using an electron saturation velocity of $2 \times 10^7$ \si{\centi \meter \per \second} and a drain-source current $I_{DS}$ $\sim$ 75 mA mm$^{-1}$, a typical value at the low-noise bias \cite{schleeh2013cryogenic}.

\begin{figure}
    \centering
    \includegraphics[width=0.6
    \linewidth]{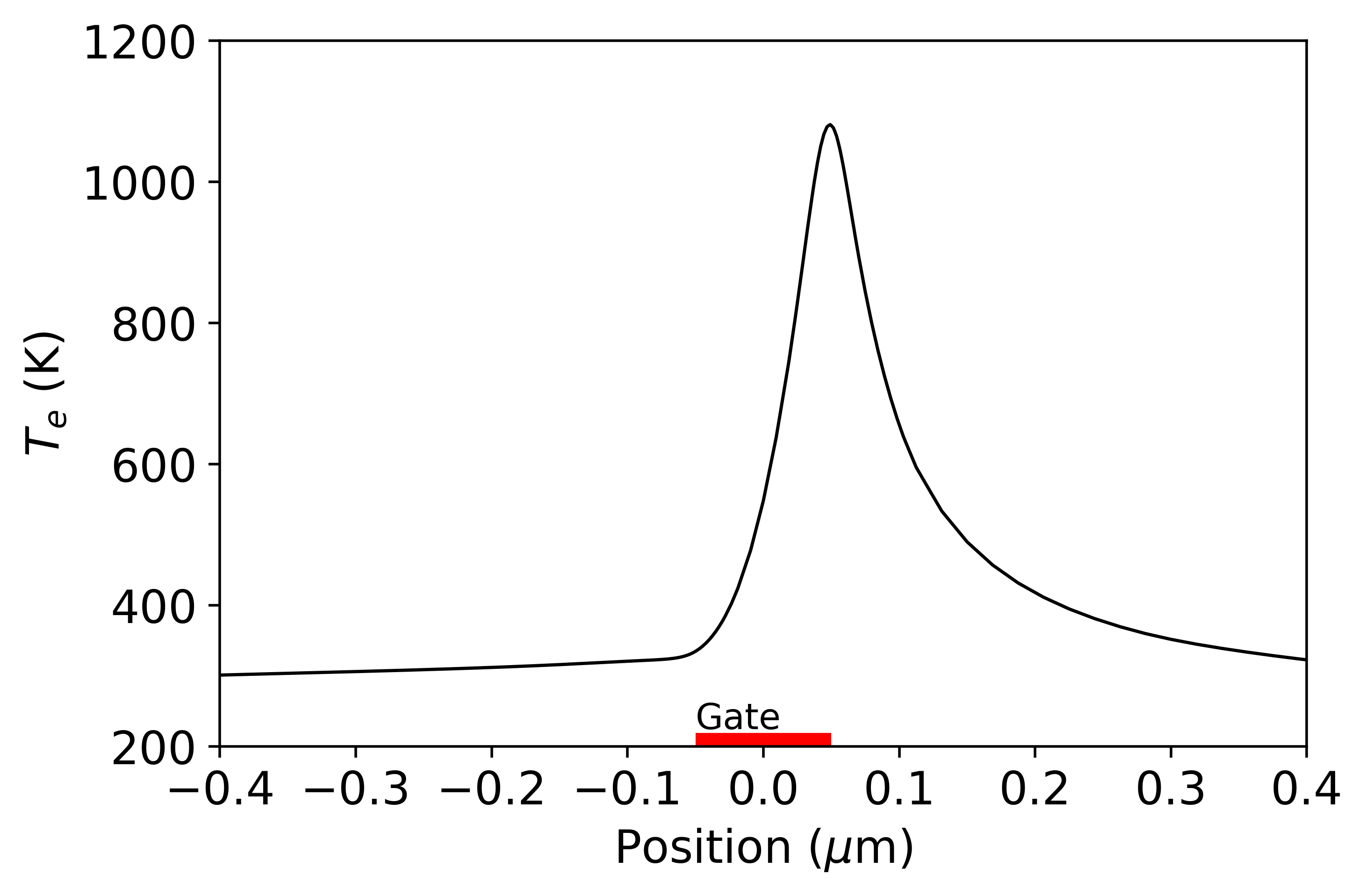}
    \caption{Electron temperature $T_e$ versus position along the channel computed using Sentaurus TCAD for $V_{DS} = 0.5$ V, $V_{GS} = 0$ V. $T_e$ peaks at the drain edge of the gate and decreases towards the drain as electrons lose energy by optical phonon emission.
    }
    \label{fig:etempfromTCAD}
\end{figure}

The hot electron fraction $\eta$ is determined by the conduction band offset $\Delta E_c$ and $V_{GS}$ for a given electron temperature $T_e$. For non-degenerate electrons in the saturated region under the gate, $\eta$ can be determined using Boltzmann statistics as $\eta = \exp(-(\Delta E_c - V_{GS})/k_BT_e)$. Therefore, to compute $\eta$, we must specify $T_e$ in the channel. We obtained a numerical estimate of its magnitude for a 100 nm gate InAlAs/InGaAs HEMT with $V_{DS} = 0.5$ V, $V_{GS} = 0$ V at a lattice temperature of 300 K  using Synopsys TCAD to solve the hydrodynamic and Poisson equations in a provided template structure \cite{synopsys}. The result is shown in Fig.~\ref{fig:etempfromTCAD}.
We observe that $T_e$  equals  the lattice temperature at the source, increases to a peak value at the drain side of the gate edge due to heating by the electric field, and decreases towards the drain as electrons lose energy by optical phonon emission. This calculation shows that peak electron temperatures in the HEMT are on the order of 1000 K around the low-noise bias point, although this value could vary by several hundred K depending on the device and bias conditions. With $T_e$ estimated, we can now compute $\eta$.  For $\Delta E_c \sim 0.5$ eV \cite{schwierz2003modern} and $T_e \sim 1000$ K, we find $\eta \sim 0.3$\%.

Using these numerical parameters, we can now use  Eq.~\ref{Td} to estimate $T_d$.   We find $T_d \sim 330$ K. This  value   is of the same order as those reported in modern HEMTs \cite{schleeh2013cryogenic, heinz2020noise}. Due to uncertainties regarding the exact values of the parameters in Eq.~\ref{Td}, we note that this value of $T_d$ should be regarded as an order of magnitude estimate rather than a precise value. However, the correct order of magnitude for the typical values of reported drain temperatures is evidence in support of the theory.

\begin{figure*}[!htb]
{
\includegraphics[width=\textwidth]{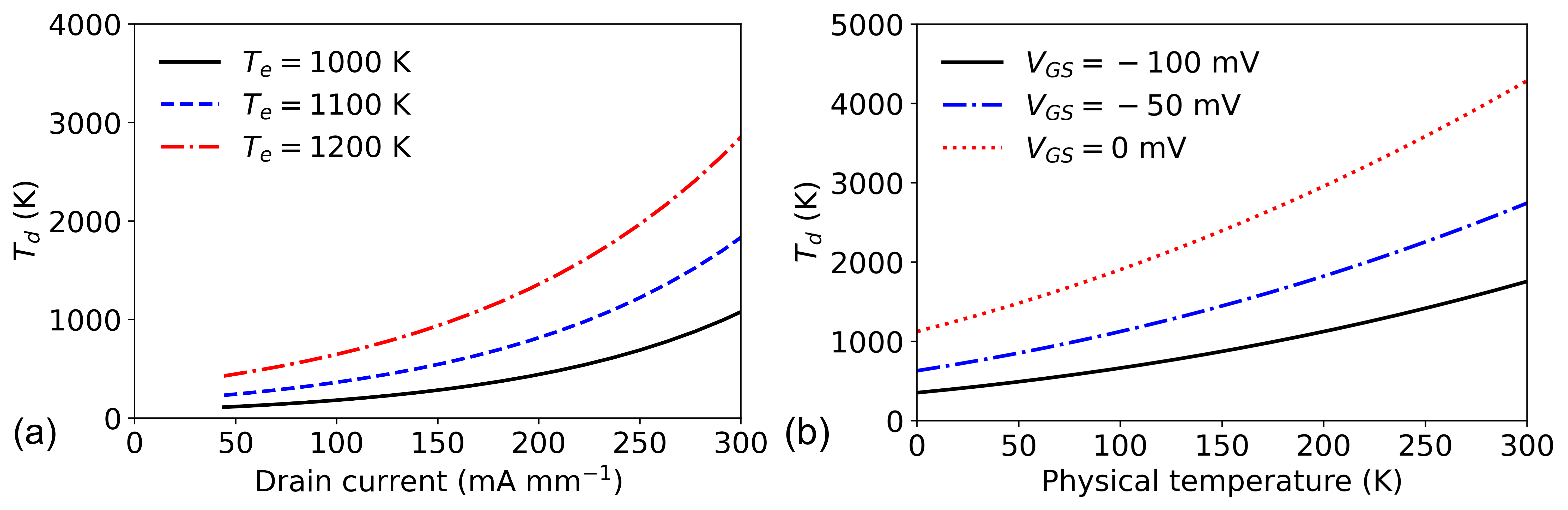}
\phantomsubcaption\label{TdIds}
\phantomsubcaption\label{TdTamb}
}
\caption{(a) Drain noise temperature $T_d$ versus $I_{DS}$. Increasing $V_{GS}$ and hence $I_{DS}$ lowers the energy barrier for thermionic emission, leading to higher $T_d$. Transfer characteristics were obtained from Fig.~4.1 of Ref.~\cite{schleeh2013cryogenic}. (b) $T_d$ versus physical temperature $T$ for $\Delta T = 1000$ K. The occupation of electronic states above the Fermi energy increases with temperature, and consequently, $T_d$ increases with $T$ as the hot electron fraction $\eta$ increases.}
\end{figure*}

We now examine the dependencies of $T_d$ predicted from Eq.~\ref{Td}. Previous works have reported a dependence of $T_d$ on $I_{DS}$ \cite{pospieszalski2005extremely, heinz2020noise} as well as physical temperature \cite{murti2000temperature,weinreb2021low,munoz1997drain}. The present theory predicts a dependence  of $T_d$ on $V_{GS}$ since $V_{GS}$ changes the Fermi level of the electrons under the gate, thus altering the population of hot electrons able to thermionically emit out of the channel as experimentally shown in Fig.~\ref{ndrmag}.

To verify that this dependence is predicted by the model, we plot $T_d$ versus $I_{DS}$  in Fig.~\ref{TdIds}. The values of $I_{DS}$ are estimated from the transfer characteristics of an InP HEMT for $V_{DS} = 0.5$ V \cite{schleeh2013cryogenic}. We observe a  dependence of $T_d$ on $I_{DS}$ which compares reasonably with experiments (see Fig.~5
of Ref. \cite{pospieszalski2005extremely}, Figs.~4 and 5 of Ref.~\cite{heinz2020noise}). In addition to qualitatively reproducing the experimental drain temperature--drain current relationship, the theory offers a physical explanation for this dependence as arising from the dependence of the hot electron fraction on Fermi level, which is controlled by $V_{GS}$.

We next examine the dependence of $T_d$ on physical temperature. Several authors have reported a temperature dependence of $T_d$, as in Fig.~8 of Ref.~\cite{murti2000temperature}, Fig.~8 of Ref.~\cite{weinreb2021low}, and Fig.~3 of Ref.~\cite{munoz1997drain}. On the other hand, other noise measurements were reported to be consistent with a temperature-independent $T_d$ \cite{pospieszalski2017dependence}.
Figure~\ref{TdTamb} illustrates how a dependence of $T_d$ on physical temperature may arise based on RST. For a non-degenerate electron gas, the electronic heat capacity is constant \cite{kittel1998thermal} so that  $T_e = T + \Delta T$ where $\Delta T$ denotes the electron temperature increase and is independent of $T$. In this figure, $\Delta T$ was chosen as 1000 K so that the computed $T_d$ is  consistent with reported cryogenic values in modern HEMTs (see Fig.~10 of Ref.~\cite{schleeh2012characterization}). The figure shows that $T_d$ can vary with physical temperature because $T_e$ varies linearly with physical temperature, which in turn affects $T_d$ through $\eta$. We observe that this dependence is more pronounced at higher physical temperatures, the parameter range studied in Ref. \cite{weinreb2021low}. At lower temperatures below 100 K, the dependence is weaker and may be more difficult to discern experimentally relative to room temperature measurements considering the challenge of accurately extracting  the drain temperature from microwave noise data. The weaker dependence below 100 K may also account for the conclusions of Ref.~\cite{pospieszalski2017dependence} that $T_d$ is independent of temperature because that study did not consider temperatures above 100 K. A quantitative comparison of the calculated dependence to experiment is difficult because $T_d$ data are often reported with $I_{DS}$ held constant, requiring shifts in $V_{GS}$ to compensate for changes in mobility, conduction band offset, and related quantities with temperature. Such changes are not included in the present calculation.


\section{Discussion}

We have presented evidence that drain noise in HEMTs can be attributed to the partition noise arising from RST of electrons from the channel to the barrier. We now discuss this finding in the context of prior explanations of drain noise. The first explanation for drain noise in the saturated region was due to Pucel \etal \cite{pucel1975signal}, who described the noise current in terms of the generation of dipole layers formed by random electron scattering events. However, their theory did not make testable predictions and so obtaining evidence to support it is difficult. Other authors have attributed noise in GaAs FETs \cite{frey1976effects, baechtold1972noise} and Si MOSFETs \cite{hendriks1988diffusion} to intervalley scattering. However, in modern InP HEMTs the $\Gamma - L$ separation  in the In$_{x}$Ga$_{1-x}$As channel (0.55 eV at 300 K and $x = 0.53$ \cite{cheng1982measurement}) exceeds the conduction band offset so that RST is expected to occur prior to intervalley transfer. Experimental evidence for this expectation has been reported in AlGaAs/GaAs heterostructures, where noise at intermediate fields---below the threshold for intervalley transfer---is attributed to RST \cite{aninkevivcius1993comparative}.

Recently,  drain noise has been suggested to arise from  a suppressed shot noise mechanism in which electrons travel nearly ballistically from source to drain \cite{pospieszalski2017limits}. However, this explanation is inconsistent with conclusions drawn from extensive studies of 2DEGs in semiconductor quantum wells. Time-resolved differential transmission spectra indicate that photo-excited electrons thermalize within around 200 fs, implying that electron-electron scattering is several times faster than this timescale \cite{knox1986femtosecond}. This relaxation time ($\sim$ 100s of fs) corresponds to a mean free path of tens of nm, which is much smaller than the $\sim$ 1 $\mu$m source-drain spacing. Further, hot electrons lose energy to the lattice on the drain side of the gate by optical phonon emission \cite{shah1978hot}, further disrupting ballistic transport across the channel.
These results indicate that electrons experience a non-negligible number of scattering events as they transit from source to drain, a finding that is incompatible with shot noise.


The present theory is able to account for the reported magnitude and trends of drain noise and is consistent with the known properties of 2DEGs. Further, it makes testable predictions that can be verified by experiment. For instance, prior studies of RST indicate that the transmission probability $\gamma$ may depend on the electronic structure of the barrier \cite{bigelow1991observation}. This finding implies that the drain noise arising from RST should depend to some degree on properties of the barrier such as the spacer thickness. Such a dependence was indeed observed in a recent study that found the microwave noise of InP HEMTs to be affected by the spacer thickness, although further analysis is required to verify that this change in noise can be primarily attributed to $T_d$ \cite{li2021red}.  This prediction of the theory is also consistent with prior studies that observed a dependence of noise on the shape of the quantum well, where the shape was controlled by channel structure and composition \cite{matulionis1997qw}.



Finally, we consider the predictions of the theory regarding how drain noise may be suppressed. A large $\Delta E_c$ is desired in HEMTs to maximize the channel sheet density \cite{schwierz2003modern}. Our theory predicts that $\Delta E_c$ is also important to suppress RST and hence drain noise. Minimizing RST requires increasing $\Delta E_c/k_B T_e$ so that the hot electron fraction decreases. A lower $T_e$ can be achieved by decreasing the InAs content of the channel and hence increasing the effective mass, but this change must be balanced against the need for high mobility and hence higher InAs content. An increase in conduction band offset, on the other hand, can be achieved without affecting the channel by reducing the InAs mole fraction of the barrier. Studies of HEMTs with barrier composition (InAs)$_x$, $0.3 < x < 0.5$, reported decreased RST in devices for smaller $x$ \cite{bahl1991strained}. However, $x$ must be chosen accounting for the lattice mismatch between the channel and barrier that can lead to the formation of misfit dislocations that negatively impact the noise.

\begin{figure}
    \centering
    \includegraphics[width=0.6
    \linewidth]{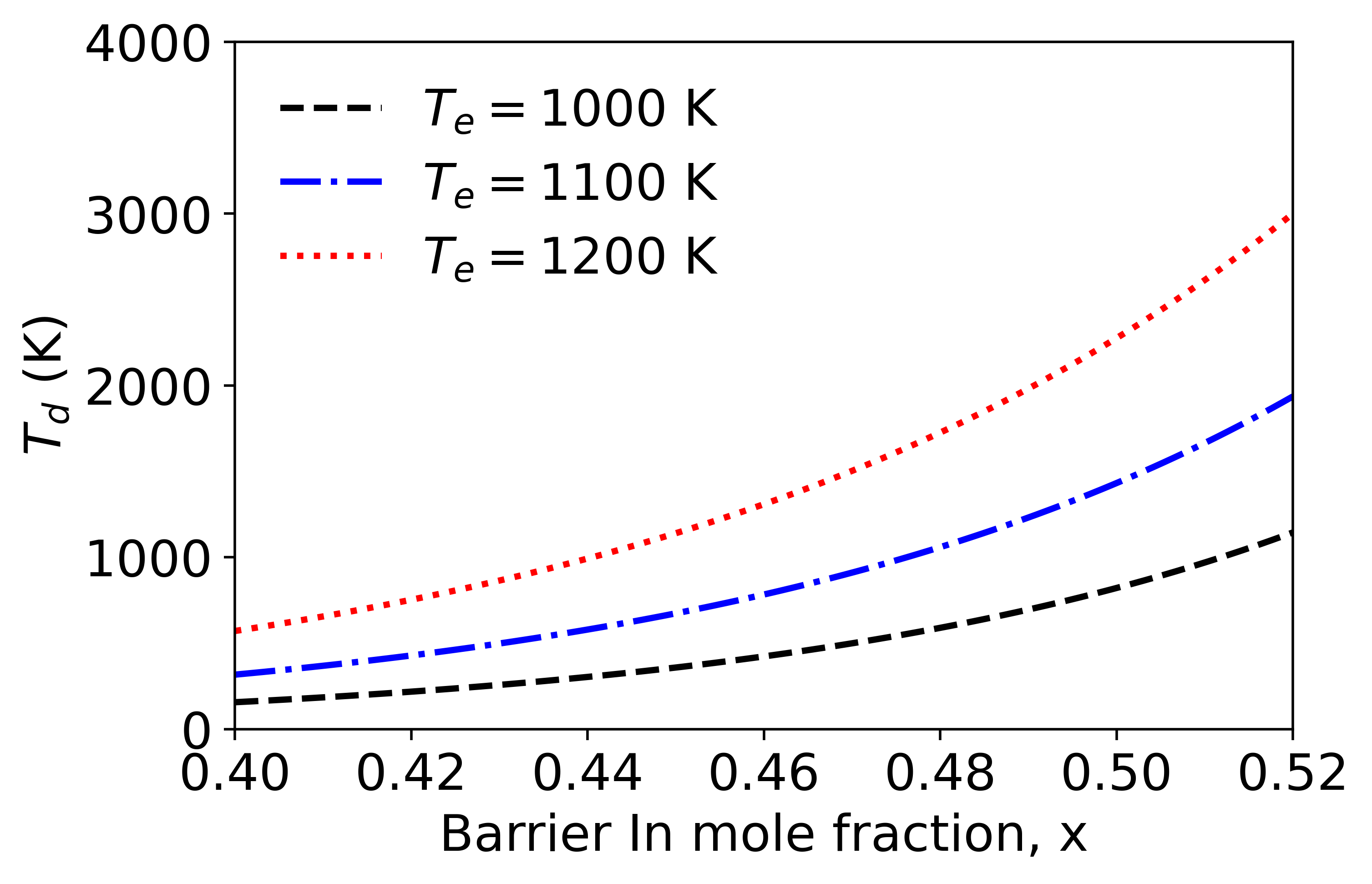}
    \caption{Drain temperature versus barrier mole fraction, $x$ in In$_x$Al$_{1-x}$As. A reduction in $T_d$ is observed as $x$  is decreased due to an increase in $\Delta E_c$.
    }
    \label{fig:Tdx}
\end{figure}

We quantify the impact of varied InAs mole fraction in the barrier on $T_d$ by  obtaining $\Delta E_c$ for each $x$ \cite{bahl1991strained}, using these values to calculate the sheet density in the barrier, and following the same analysis as described in Sec.~\ref{results}. The result is shown in Fig.~\ref{fig:Tdx}. We observe a marked decrease in $T_d$ as $x$ is reduced from its lattice-matched value of 0.52 to 0.46, followed by a slower decrease from 0.46 to 0.4. Following Pospieszalski's noise model \cite{pospieszalski1989modeling}, a factor of $\sim$ 2 reduction in $T_d$ as seen when $x$ changes from 0.52 to 0.46 translates to a factor of 1.4 reduction in the minimum noise temperature. This analysis suggests that further improvements to the noise figure of HEMTs can be realized by optimizing the barrier InAs mole fraction. 


\section{Summary}
We have reported a theory of drain noise in  high electron mobility transistors based on microwave partition noise arising from real-space transfer of electrons from the channel to the barrier. The theory successfully explains the reported magnitude and dependencies of $T_d$. The theory predicts that $T_d$ can be decreased by altering the barrier composition to increase the conduction band offset and thus decrease the occurrence of RST. Our results may  guide the design of HEMTs with lower microwave noise figure.

\section{acknowledgements}
The authors thank Jan Grahn and Junjie Li at Chalmers University of Technology for useful discussions and providing the data shown in Figure~\ref{IV}. I.E. was supported by the National Science Foundation Graduate Research Fellowship under Grant No. DGE-1745301. A.Y.C. and A.J.M. were supported by the National Science Foundation under Grant No.~1911220.  Any opinions, findings, and  conclusions or recommendations expressed in this material are those of the authors and do not
necessarily reflect the views of the National Science Foundation.

\clearpage

\bibliographystyle{is-unsrt}
\bibliography{bib}

\end{document}